\documentclass[journal,twoside,web]{ieeecolor}
\usepackage{tmi}
\usepackage{cite}
\usepackage{amsmath,amssymb,amsfonts}
\usepackage{algorithmic}
\usepackage{graphicx}
\usepackage{textcomp}
\usepackage{booktabs}

\def\BibTeX{{\rm B\kern-.05em{\sc i\kern-.025em b}\kern-.08em
    T\kern-.1667em\lower.7ex\hbox{E}\kern-.125emX}}
\begin{document}
\title{AS-Mamba: Asymmetric Self-Guided Mamba Decoupled Iterative Network for Metal Artifact Reduction}

\author{Bowen Ning, Zekun Zhou, Xinyi Zhong, Zhongzhen Wang, HongXin Wu, HaiTao Wang, Liu Shi, and Qiegen Liu, \IEEEmembership{Senior Member, IEEE}
	\thanks{This study was funded by National Natural Science Foundation of China (621220033), Early Stage Young Scientific and Technological Talent Training Foundation of Jiangxi Province (Grant: 20252BEJ730005), and Nanchang University Youth Talent Training Innovation Fund Project (Grant: XX202506030012). This work was supported by data from LargeV Instrument Corp., Ltd. (Beijing, China).}%
	\thanks{B. Ning and Z. Zhou contributed equally to this work. (Corresponding authors: Liu Shi and Qiegen Liu.) B. Ning, X. Zhong, L. Shi, and Q. Liu are with the School of Information Engineering, Nanchang University, Nanchang 330031, China (e-mail: 416100240210@email.ncu.edu.cn; 416100250217@email.ncu.edu.cn; shiliu@ncu.edu.cn; liuqiegen@ncu.edu.cn). Z. Zhou is with the School of Mathematics and Computer Sciences, Nanchang University, Nanchang 330031, China (e-mail: ZeKunZhou@email.ncu.edu.cn). H. Wang is with The First Affiliated Hospital, Jiangxi Medical College, Nanchang University, Nanchang 330006, China (e-mail: wanghaitao20000103@163.com). Z. Wang and H. Wu are with LargeV Instrument Corp., Ltd., Beijing 100084, China (e-mail: wangzhongzhen@largev.com; wuhongxin@largev.com).}
}
\maketitle

\begin{abstract}
Metal artifact significantly degrades Computed Tomography (CT) image quality, impeding accurate clinical diagnosis. However, existing deep learning approaches, such as  CNN and Transformer, often fail to explicitly capture the directional geometric features of artifacts, leading to compromised structural restoration. To address these limitations, we propose the Asymmetric Self-guided Mamba (AS-Mamba) for metal artifact reduction. Specifically, the linear propagation of metal-induced streak artifacts aligns well with the sequential modeling capability of State Space Models (SSMs). Consequently, the Mamba architecture is leveraged to explicitly capture and suppress these directional artifacts. Simultaneously, a frequency-domain correction mechanism is incorporated to rectify the global amplitude spectrum, thereby mitigating intensity inhomogeneity caused by beam hardening. Furthermore, to bridge the distribution gap across diverse clinical scenarios, we introduce a self-guided contrastive regularization strategy. Extensive experiments on public and clinical dental CBCT datasets demonstrate that AS-Mamba achieves superior performance in suppressing directional streaks and preserving structural details, validating the effectiveness of integrating physical geometric priors into deep network design.
\end{abstract}


\section{Introduction}
\label{sec:introduction}
\normalcolor
\IEEEPARstart{C}{OMPUTED }tomography (CT) is crucial in modern clinical practice, offering high-resolution cross-sectional images for diagnosis and treatment planning. However, its utility is often compromised by metallic implants like dental fillings and prosthetics, which cause severe beam hardening and photon starvation artifacts. These artifacts degrade image quality, obscuring anatomical structures and potentially leading to diagnostic errors or inaccurate treatment planning \cite{b1, b2, b3}. Consequently, developing effective MAR techniques remains a vital clinical challenge, particularly for modalities heavily impacted by dental hardware, such as dental Cone Beam CT.
\begin{figure}[!t]
	\centerline{\includegraphics[width=\linewidth]{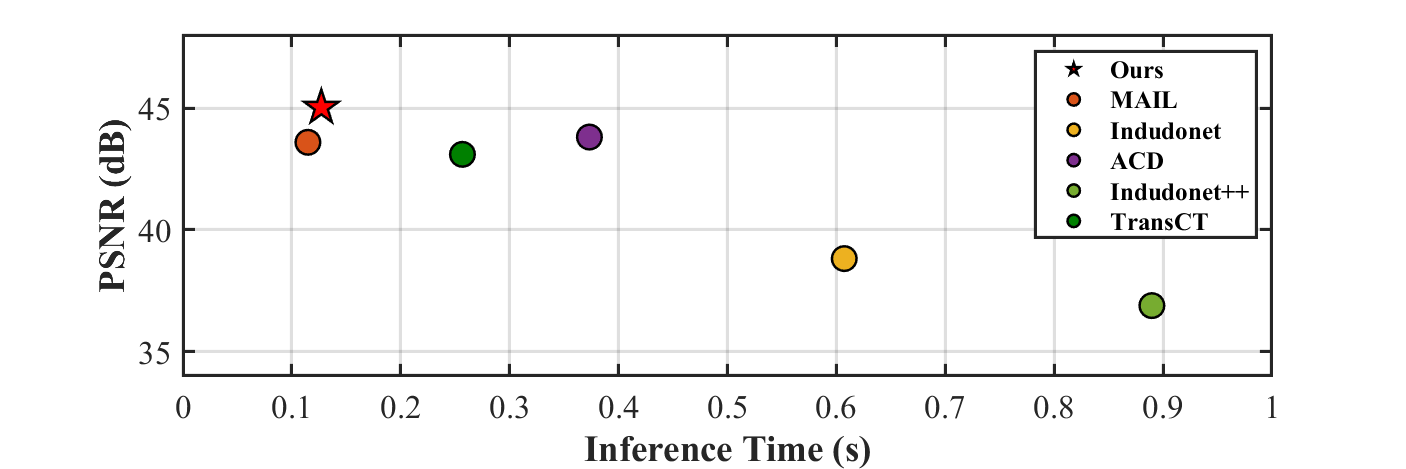}}
	\caption{Comparison of inference speed and image quality on the
		DeepLesion dataset. The x-axis represents the average inference time
		per image, and the y-axis denotes the PSNR(dB) value. Our method achieves superior reconstruction quality
		with highly competitive inference speed compared to other state-of-the
		arts. }
	\label{fig1}
\end{figure}

Early sinogram-based interpolation methods, such as LI \cite{b4} and NMAR \cite{b5}, treat metal traces as missing data. However, relying on local interpolation often disrupts global geometric continuity, inevitably causing secondary artifacts in complex anatomical scenarios. Alternatively, iterative reconstruction methods \cite{b6, b34} incorporate physical priors to mitigate artifacts. Nevertheless, they struggle to accurately model the complex non-linear physics induced by high-density metal, and their prohibitive computational costs further hinder clinical adoption.

The advent of deep learning has driven significant advancements in MAR \cite{b7, b8}. Early CNNs typically utilize encoder-decoder architectures to learn the mapping from corrupted to clean images \cite{b8, b9, b10}. To integrate physical priors, Lin et al. \cite{b18} proposed a dual-domain learning framework that links the sinogram and image domains through a Radon inversion layer to simultaneously restore projection data and image details. Subsequently, Wang et al. \cite{b26} developed a deep unfolding architecture that incorporates an adaptive convolutional dictionary model to iteratively decompose metal artifacts into structural and texture components via a proximal gradient algorithm. While these model-driven networks improve consistency through physical constraints, their ability to represent long-range correlated structures remains limited by the intrinsic locality of convolution. To address this, Transformers \cite{b14} were introduced to capture global dependencies. However, their patch-based processing often fragments the continuous linear trajectories of streak artifacts, and their quadratic complexity hinders high-resolution clinical imaging.

Recent advances in State Space Models (SSMs), especially the Mamba architecture [16], are gaining prominence for their linear complexity and effectiveness in capturing long-range dependencies [17]. Within MAR, Mamba’s sequential scanning mechanism aligns naturally with the linear trajectory of metal streak artifacts, allowing it to trace and suppress these directional distortions effectively. However, relying solely on spatial-domain modeling often causes feature confusion during state propagation, since both streak artifacts and anatomical details appear as high-frequency signals. This makes it difficult to separate artifacts from true edges. Inspired by frequency-domain solutions to similar problems in general vision tasks [21, 33], we argue that fully leveraging Mamba for MAR requires explicitly decoupling and regularizing frequency components during artifact modeling.

To address these challenges, we propose the Asymmetric Self-guided Mamba Decoupled Iterative Network (AS-Mamba). Unlike recent works such as MDS-MAR \cite{b36}, which use Mamba merely as a generic feature extractor, we fundamentally align the network topology with the artifact propagation mechanism. Our framework employs an asymmetric dual-branch architecture to minimize mutual interference between artifacts and anatomical structures. Specifically, metal artifacts can be characterized as a combination of non-local high-frequency streaks, constrained by linear propagation trajectories, and global low-frequency shading distortions, caused primarily by beam-hardening effects. To address these distinct characteristics, an asymmetric dual-branch modeling strategy is adopted. The high-frequency branch leverages the sequential modeling capability of Mamba to perform fine-grained suppression of directional streak artifacts along the corresponding propagation paths. Simultaneously, the low-frequency branch employs Fourier-domain modeling to rectify the global amplitude spectrum, thereby effectively mitigating intensity inhomogeneity. Furthermore, to handle the distribution shifts inherent in diverse clinical data, we introduce a Self-Guided Contrastive Regularization (SGCR) strategy within an iterative refinement loop, enhancing model generalization.

The main contributions of our work are summarized as follows:

\begin{itemize}
	\item \textbf{Physics-Informed Mamba Based Artifact Suppression.} We propose a physics-guided Mamba-based method that suppresses streak artifacts in the high-frequency domain. By leveraging the inherent alignment between the linear propagation of metal-induced X-ray streaks and Mamba’s selective scanning mechanism, our approach structurally models streak propagation in high-frequency space. This enables effective suppression of non-local, directional artifacts along their geometric trajectories.
	
	\item \textbf {Low-Frequency Amplitude Spectrum Correction}. We propose a Dual Enhancement Network (DEN) to alleviate intensity inhomogeneity caused by physical phenomena such as beam hardening. This module explicitly corrects the corrupted amplitude spectrum within the low-frequency components, thereby restoring global structural consistency.
	
	\item \textbf {Scene-Aware Contrastive Regularization}. We introduce a self-guided contrastive regularization loss to enhance the capability of the model in modeling structural consistency across diverse clinical imaging scenarios. This strategy effectively addresses the inconsistent artifact distributions caused by anatomical variations and different metallic implants. Consequently, the model maintains stable artifact suppression and structural fidelity under various clinical conditions.
\end{itemize}

The rest of this article is organized as follows: Section II introduces the theoretical preliminaries. Section III details the proposed AS-Mamba framework and the self-guided contrastive regularization strategy. Section IV presents the experimental settings and results. Section V concludes this paper.

\section{Preliminary}
\subsection{Physical  Characteristics of Metal Artifacts}
CT reconstruction typically relies on a monochromatic X-ray spectrum and linear material attenuation. However, in clinical practice, high-density metallic implants violate this linearity due to beam hardening and polychromaticity, where low-energy photons are preferentially absorbed. Consequently, the measured projection $P$ deviates from the ideal Radon transform, becoming a non-linear function of energy-dependent attenuation:
\begin{equation}
	P(\theta, s) = -\ln \int \eta(E) \exp\left(-\int_{L_{\theta,s}} f_E(x) dl\right) dE,
	\label{eq:beam_hardening}
\end{equation}
where $\eta(E)$ denotes the X-ray energy spectrum. 

This degradation is further exacerbated by photon starvation and scattering effects. Metallic objects with extreme attenuation coefficients can attenuate the X-ray beam to levels approaching the detector's noise floor. Upon logarithmic transformation, the dominance of electronic and Poisson noise introduces high-amplitude streak patterns that propagate across the image domain. 

\subsection{State Space Models and Mamba}SSMs provide a framework for modeling sequence data by mapping a 1D input sequence $x(t)$ to an output $y(t)$ through evolving latent states $h(t)$. The foundational continuous-time dynamics are governed by the following linear ordinary differential equations:
\begin{equation}
	h'(t) = Ah(t) + Bx(t), \quad y(t) = Ch(t) + Dx(t),\label{eq:continuous_ssm}
\end{equation}
where $A, B, C,$ and $D$ are matrices defining the system dynamics. To apply this continuous model to discrete signals in deep learning, it must be discretized. Using the common zero-order hold  technique with a time-step $\Delta$, the system transforms into a discrete-time recursive equations:
\begin{equation}
	h_t = \bar{A}h_{t-1} + \bar{B}x_t, \quad y_t = Ch_t + Dx_t,\label{eq:discrete_ssm}
\end{equation}
where the discrete parameters are derived from their continuous counterparts as $\bar{A}=e^{\Delta A}$ and $\bar{B}=(\Delta A)^{-1}(e^{\Delta A}-I)\Delta B$.

While traditional SSMs are time-invariant, the recently proposed Mamba architecture improves this by introducing a data-dependent selection mechanism. This allows the model parameters to vary based on the current input, enabling the network to selectively remember relevant information and forget irrelevant data along long sequences. Furthermore, Mamba employs a parallel scanning algorithm to compute this recurrence efficiently during training.

\section{Method}

We design AS-Mamba with an asymmetric dual-branch architecture and we begin by presenting the motivation behind our core design principle of asymmetric frequency-domain processing.Subsequently, we describe the overall network architecture, elaborate on the key components.

\subsection{\textcolor{subsectioncolor}{Motivation}}
Metal implants in CT imaging introduce complex artifacts primarily through beam hardening and photon starvation, which manifest as streaks and global distortions that severely degrade image quality \cite{b8}. To address these effects, Park et al. \cite{b37} provided a theoretical foundation for artifact mitigation by modeling the reconstruction problem under the assumption that beam hardening is the primary degradation source. Using microlocal analysis, they decomposed the CT image into an ideal monochromatic component and a metal artifact term, as formulated in Lemma 1, facilitating subsequent modeling and removal strategies.

\vspace{0.2cm}
\noindent \textbf{Lemma 1 \cite{b36}:}
The reconstructed CT image $f_{CT}$ can be decomposed into an ideal monochromatic term $f_{E_0}$ and a metal artifact term $f_{MA}$:
\begin{equation}
	f_{CT}(x) = f_{E_0}(x) + f_{MA}(x),
\end{equation}
where the artifact term $f_{MA}$ is explicitly defined as the back-projection of the nonlinear beam hardening error:
\begin{equation}
	f_{MA}(x) = -\frac{1}{4\pi}\mathcal{R}^*\mathcal{I}^{-1} \left[ \ln \left( \frac{\sinh(\eta \mathcal{R}\chi_D)}{\eta \mathcal{R}\chi_D} \right) \right](x),
\end{equation}
where $\mathcal{R}^*$ and $\mathcal{I}^{-1}$ denote the back-projection operator and the inverse Hilbert transform, respectively. The parameter $\eta$ represents the beam hardening coefficient, and $\mathcal{R}\chi_D$ corresponds to the projected thickness of the metallic implants.
\vspace{0.2cm}

\begin{figure}[t!]
	\centering
	
	\includegraphics[width=1\columnwidth]{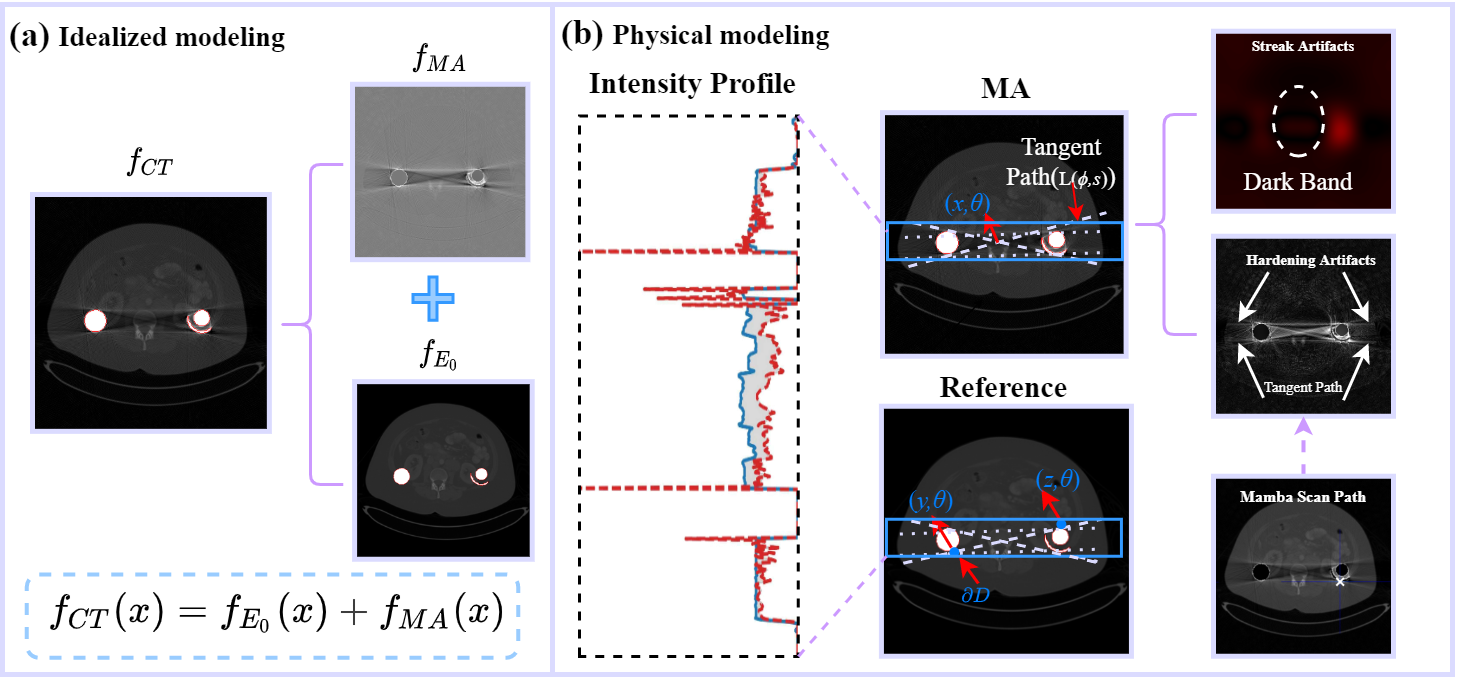}
	
	\caption{Analysis of metal artifact mechanisms. (a) Physical decomposition of the corrupted image $f_{CT}$ into the artifact term $f_{MA}$ and ideal term $f_{E_0}$. (b) Mechanism dissection: The intensity profile (left) quantifies the low-frequency hardening distortion, where blue and red lines represent the Reference and MA, respectively. The schematic (right) illustrates the anisotropy of high-frequency streaks, where singularities from boundary points $y, z \in \partial D$ propagate along linear trajectories that structurally align with the Mamba scan path.}
	\label{fig:wave_analysis} 
	\vspace{-0.5cm}
\end{figure}

Visual insights into metal artifacts are provided in Fig. \ref{fig:wave_analysis}. Specifically, Fig. \ref{fig:wave_analysis}(a) depicts the decomposition of a metal-corrupted CT image $f_{CT}$ into its ideal monochromatic component $f_{E_0}$ and the associated metal artifact term $f_{MA}$. The directional characteristics of streak and beam-hardening artifacts are further illustrated in Fig. \ref{fig:wave_analysis}(b). In the context of metal artifacts, the hardening component primarily manifests as low-frequency intensity distortion. In contrast, the streak component exhibits pronounced anisotropy, with its morphology dominated by sharp directional singularities aligned with geometric tangent lines. Such a streak singularity typically arises when the projection data satisfy the following dimensional condition:
\begin{equation}
	\dim(\text{Span}[\Sigma_{(\varphi, s)}(\mathcal{R}\chi_{D})]) = 2,
\end{equation}
where $\Sigma$ denotes the singularity set in the phase space. These anisotropic streaks propagate along linear trajectories, and the structure of the Mamba scan path is highly aligned with their propagation directions.

Based on the above theoretical insights and the physical origins of the artifacts, we adopt corresponding strategies to effectively suppress streak and hardening artifacts. Specifically, since streak artifacts propagate linearly along specific geometric paths, the Mamba scan path is highly aligned with their propagation directions, thereby fully leveraging the network’s ability to capture non-local streaks for effective suppression. Meanwhile, to address the global shading artifacts caused by nonlinear beam hardening, we employ DEN to rectify the global amplitude spectrum and restore the intensity homogeneity of the anatomical background.

\subsection{\textcolor{subsectioncolor}{Non-Local Streak Suppression Model}}
\normalcolor
The input to our framework is a composite tensor $X_{i}$, constructed by concatenating the metal-artifact-corrupted image $X_{m}$, which preserves high-fidelity data details, with the linear interpolation-based prior image $X_{L}$, which provides essential low-frequency structural guidance, along the channel dimension.
We employ a single-level 2D Discrete Wavelet Transform (DWT) to decompose $X_{i}$ into frequency sub-bands.
The decomposition at spatial location $(i, j)$ for a specific sub-band $s \in \{LL, LH, HL, HH\}$ is computed via separable convolution and downsampling:
\begin{equation}
	X_{s}[i, j] = \sum_{m,n} \phi_{r}[m] \phi_{c}[n] \cdot X_{i}[2i-m, 2j-n],
	\label{eq:dwt_unified}
\end{equation}
where $\phi_{r}, \phi_{c} \in \{g, h\}$ denote the 1D scaling and wavelet filters applied along the row and column dimensions, respectively.
For instance, the approximation band $X_{L}$ corresponds to $\phi_{r}=g, \phi_{c}=g$, while the diagonal detail band $X_{HH}$ corresponds to $\phi_{r}=h, \phi_{c}=h$.

\begin{figure*}[t!]
	\centerline{\includegraphics[width=\textwidth]{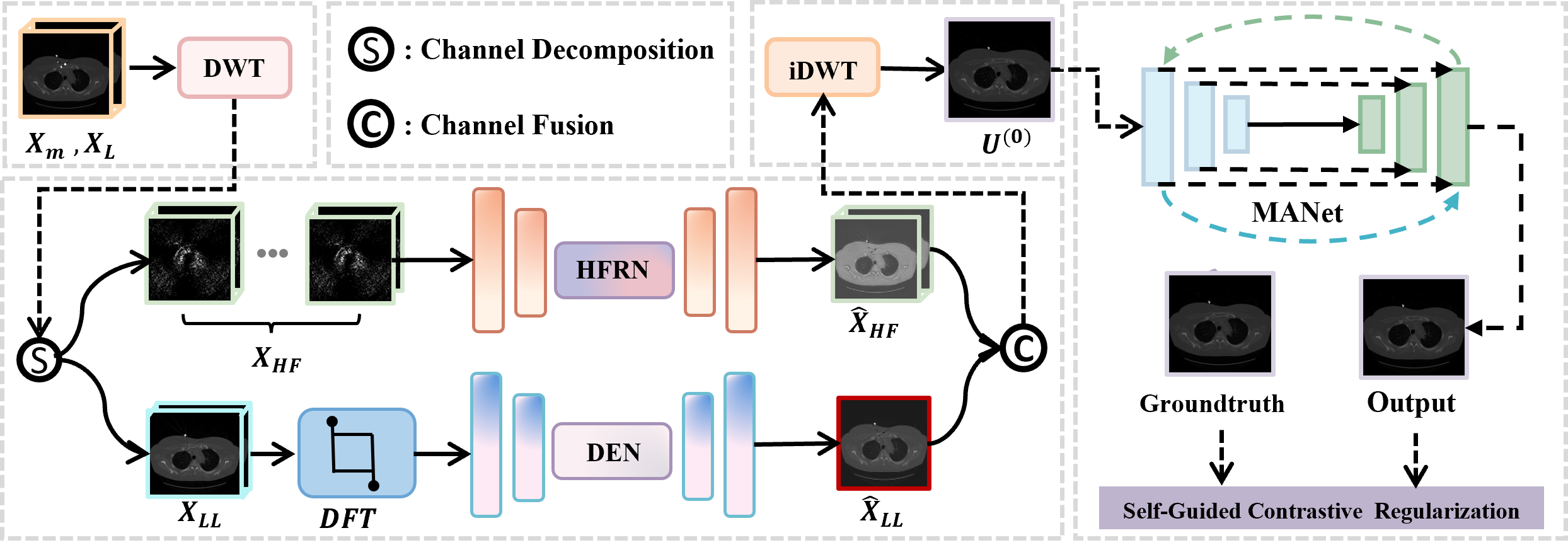}}
	\caption{Overall framework of AS-Mamba, consisting of a frequency-decoupled reconstruction module and an iterative refinement module: 
		(1) The frequency-decoupled module targets specific artifact components.
		The High Frequency Reduction Network (HFRN) utilizes Mamba to suppress directional streaks in the high-frequency domain \textnormal{$X_{HF}$} , while the Dual Enhancement Network (DEN) corrects beam hardening in the low-frequency domain \textnormal{$X_{LL}$}.
		(2) The iterative module ensures global structural consistency. The Iterative MANet progressively refines the initial reconstruction \textnormal{$U^{(0)}$} to recover anatomical details, enhanced by Self-guided Contrastive Regularization.}
	\label{fig:framework}
	\vspace{-0.5cm}
\end{figure*}

The fundamental unit, MambaBlock ($\mathcal{MB}$), utilizes a dual-residual structure to capture long-range dependencies. It first processes the layer-normalized input via a selective state space model (SSM) branch, where spatial features are flattened to trace directional streaks. The intermediate feature map $F_{m}$ is computed as:
\begin{equation}
	F_{m} = F_{in} + \alpha \cdot \text{SSM}_{s}(\text{LN}(F_{in})).
	\label{eq:mamba_stage1}
\end{equation}
Subsequently, a Feed-Forward Network (FFN) refines this feature map to produce the final block output:
\begin{equation}
	\mathcal{MB}(F_{in}) = F_{m} + \beta \cdot \text{FFN}(\text{LN}(F_{m})),
	\label{eq:mamba_stage2}
\end{equation}
where $\text{LN}(\cdot)$ denotes Layer Normalization, and $\alpha, \beta$ are learnable scalar parameters. This design effectively models non-local artifact propagation with linear complexity.

The HFRN processes the concatenated high-frequency sub-bands, $X_{HF} = \text{Concat}(X_{LH}, X_{HL}, X_{HH})$.
The HFRN has a U-Net architecture built with MambaBlocks and employs Haar Wavelet Downsampling (HWD) in the encoder.
Let $F_{e}^{(0)} = \text{Conv}_{in}(H_{hl})$ be the initial feature maps. The encoding path consists of $N$ stages.
At stage $i$ ($i=1, \dots, N$), features are processed by an encoder MambaBlock, $\mathcal{MB}_E^{(i)}$, and downsampled via HWD:
\begin{equation}
	F_{e}^{(i)} = \text{HWD}(\mathcal{MB}_E^{(i)}(F_{e}^{(i-1)})).
\end{equation}
The bottleneck features are processed as $F_{b} = \mathcal{MB}_B(F_{e}^{(N)})$. The decoding path reconstructs features from stage $i = N-1$ to $0$.
Deeper features are upsampled, concatenated with corresponding encoder features via skip connections, and processed by a decoder MambaBlock, $\mathcal{MB}_D^{(i)}$:
\begin{equation}
	F_{d}^{(i)} = \mathcal{MB}_D^{(i)}(\text{Concat}(\text{Up}(F_{d}^{(i+1)}), F_{e}^{(i)})),
\end{equation}
where $F_{d}^{(N)} = F_{b}$ is the starting condition, and $\text{Up}(\cdot)$ is bilinear upsampling.
The final refined high-frequency output, $\hat{H}_{hl}$, is generated by a terminal convolution layer:
\begin{equation}
	\hat{X}_{HF} = \text{Conv}_{out}(F_{d}^{(0)}).
\end{equation}

\begin{figure*}[!t]
	\centerline{\includegraphics[width=\textwidth]{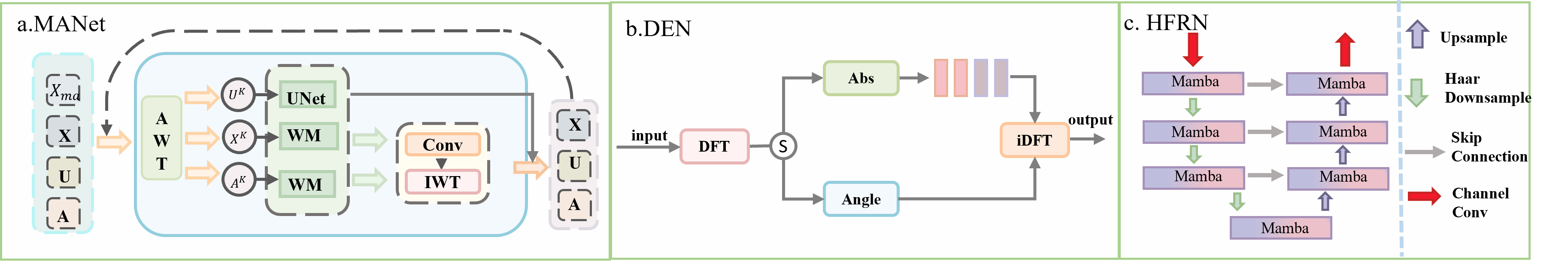}}
	\caption{Detailed architecture of the proposed core sub-modules. (a) The Iterative MANet focuses on progressive image refinement. (b) The DEN operates in the Fourier domain to rectify the amplitude spectrum.
		(c) The HFRN  designed to capture long-range dependencies for suppressing directional streak artifacts.}
	\label{fig:submodules}
\end{figure*}

\subsection{\textcolor{subsectioncolor}{Amplitude-Guided Low-Frequency Correction}}

This branch employs the DEN module to model the restoration of the Fourier domain, in order to alleviate the problem of frequency domain information loss caused by the metal hardening effect.
Let $X_{LL}$ denote the input feature map in the spatial domain with dimensions $H \times W$.
First, we compute its frequency representation $Z$ using the 2D Real Fast Fourier Transform.
The transformation for a frequency coordinate $(u, v)$ is explicitly defined as:
\begin{equation}
	Z(u, v) = \sum_{h=0}^{H-1} \sum_{w=0}^{W-1} X_{LL}(h, w) \cdot e^{-j2\pi(\frac{uh}{H} + \frac{vw}{W})}.
	\label{eq:rfft_def}
\end{equation}
This complex-valued representation is then decomposed into its amplitude spectrum $\mathcal{A} = |Z|$ and phase spectrum $\mathcal{P} = \angle Z$.
The restoration process specifically targets the amplitude spectrum $\mathcal{A}$, which encodes the structural intensity information often degraded by photon starvation.
The amplitude is enhanced additively by a learned parameter $f_{e}$, which represents a convolutional filter applied in the feature domain.
Mathematically, this operation is formulated as learning a residual function ${g}(\cdot)$ applied to the amplitude:
\begin{equation}
	\begin{split}
		\hat{\mathcal{A}} &= \mathcal{A} + \mathcal{F}\left(f_{e} \ast \mathcal{F}^{-1}(\mathcal{A})\right) \\
		&\approx \mathcal{A} + {g}(\mathcal{A}),
	\end{split}
	\label{eq:mag_enhance}
\end{equation}
where $\ast$ denotes the convolution operation, ${g}(\cdot)$ represents the parameterized enhancement network, and $\hat{\mathcal{A}}$ denotes the enhanced amplitude.
Crucially, the original phase $\mathcal{P}$ remains preserved during this process to maintain the correct spatial localization of anatomical boundaries.
Finally, we reconstruct the refined complex frequency representation $\hat{Z}$ by recombining the enhanced amplitude with the original phase:
\begin{equation}
	\hat{Z}(u, v) = \hat{\mathcal{A}}(u, v) \cdot e^{j\mathcal{P}(u, v)}.
\end{equation}
Subsequently, the spatial feature map $\hat{X}_{LL}$ is recovered via the inverse transform:
\begin{equation}
	\hat{X}_{LL}(h, w) = \frac{1}{HW} \sum_{u=0}^{H-1} \sum_{v=0}^{W-1} \hat{Z}(u, v) \cdot e^{j2\pi(\frac{uh}{H} + \frac{vw}{W})}.
	\label{eq:irfft_def_cn}
\end{equation}
This process explicitly targets the restoration of structural information encoded in the low-frequency magnitude spectrum.
The DEN cascades three such enhancement stages with convolutional groups and feature fusion to progressively refine the low-frequency features.

\subsection{\textcolor{subsectioncolor}{Iterative Correction Network}}
\normalcolor
Following specialized frequency-domain processing, the refined low-frequency $\hat{X}_{LL}$ and suppressed high-frequency $\hat{X}_{HF}$ features are synthesized via the iDWT. This produces an initial coarse reconstruction $U^{(0)} = \text{iDWT}(\hat{X}_{LL}, \hat{X}_{HF})$, used to initialize the iterative refinement. 

To effectively remove residual artifacts and ensure global structural consistency, we frame the MAR task as a variational optimization problem. Our goal is to decompose the metal-corrupted image $Y$ into a clean image $X$ and a metal artifact component $A$ within the non-metal region defined by mask $I$. Introducing an auxiliary variable $U$ to link the wavelet and spatial domains, the objective function is:

\begin{equation}
	\min_{X,A,U} | I \odot (X + A - Y) |_F^2 + \mathbf{\mu} | I \odot X - U |_F^2 + \mathcal{R}(X, A, U),
	\label{eq:objective_function}
\end{equation}
where $\mathbf{\mu}$ is the penalty parameter, and $\mathcal{R}(X,A,U)$ denotes the regularization terms.
\begin{equation}
	\mathcal{R}(X, A, U) = \lambda_1 f_1(W(I \odot A)) + \lambda_2 f_2(W(I \odot X)) + \lambda_3 f_3(U),
	\label{eq:regularization_term}
\end{equation}

with non-negative hyperparameters $\lambda_1, \lambda_2, \lambda_3$ balancing the terms, and $f_1(\cdot)$, $f_2(\cdot)$, $f_3(\cdot)$ representing specific regularizers applied to each component.

To solve \eqref{eq:objective_function}, we employ an iterative MANet, a deep unrolling network that maps the Proximal Gradient Descent steps into a cascaded structure. The network alternately updates the artifact $A$, clean image $X$, and final reconstruction $U$ across $T$ stages.

At the $k$-th stage, we first update the artifact estimate in the wavelet domain. From the data fidelity gradient, an intermediate artifact variable is computed:
\begin{equation}
	\tilde{A}_w^{(k)} = (1 - 2\mathbf{\tau_1}) A_w^{(k-1)} + 2\mathbf{\tau_1} W(Y - X^{(k-1)}),
	\label{eq:Amid}
\end{equation}
where $W(\cdot)$ is the wavelet transform, $A_w$ denotes the artifact wavelet coefficients, and $\mathbf{\tau_1}$ is a learnable step size.

This estimate is then refined by the WM module $\text{WM}_A(\cdot)$, which replaces the conventional proximal operator. The module uses Mamba’s global modeling to suppress residual artifacts, giving the updated component:
\begin{equation}
	A_w^{(k)} = \text{WM}_A(\tilde{A}_w^{(k)}).
\end{equation}

Subsequently, we update the image component $X$ by incorporating constraints from both the observation $Y$ and the auxiliary variable $U$.
The intermediate image estimate $\tilde{X}_w^{(k)}$ is calculated via a gradient step:
\begin{equation} \tilde{X}_w^{(k)} = (1 - 2\mathbf{\tau_2}) X_w^{(k-1)} + \frac{2\mathbf{\tau_2}}{\mathbf{\gamma+\delta}} \Phi^{(k)}, \label{eq:X_mid} \end{equation}

where $\Phi^{(k)}$ represents the composite guidance term derived from data fidelity and spatial consistency, defined as:
\begin{equation}
	\Phi^{(k)} = \mathbf{\gamma} W(Y - A^{(k)}) + \mathbf{\delta} W(U^{(k-1)}),
	\label{eq:X_mid_phi}
\end{equation}
where  $\mathbf{\tau_2, \gamma, \delta}$ are learnable parameters.

This noisy intermediate estimate is then processed by a dedicated image-refinement WM module, $\mathcal{WM}_X(\cdot)$, which acts as a learnable denoiser to recover structural details, resulting in the updated image coefficients $X_w^{(k)} = \mathcal{WM}_X(\tilde{X}_w^{(k)})$.

Finally, the updated wavelet coefficients $X_w^{(k)}$ are transformed back to the spatial domain via iDWT to update the final reconstruction variable $U$.
This step ensures spatial consistency and mitigates blocking effects from domain transformations:
\begin{equation}
	U^{(k)} = (1 - 2\mathbf{\tau_3}) U^{(k-1)} + 2\mathbf{\tau_3} \text{iDWT}(X_w^{(k)}),
	\label{eq:U_update}
\end{equation}
where $\mathbf{\tau_3}$ is a learnable step size governing the update rate of the final reconstruction variable $U$.
By cascading these operations for $T$ iterations with adaptively learned parameters, the iterative MANet progressively converges to a high-quality, artifact-free image $U^{(T)}$.

% =====================================================================
% E. Self-Guided Contrastive Regularization
% =====================================================================
\subsection{\textcolor{subsectioncolor}{Self-Guided Contrastive Regularization}}
To enhance perceptual fidelity, we introduce the SGCR loss, $\mathcal{L}_{S}$. We construct a contrastive triplet with the final output $U^{(T)}$ as the anchor, the ground truth $X_{gt}$ as the positive sample, and the initial coarse reconstruction  $U^{(0)}$ as the hard negative sample. By treating $U^{(0)}$ as a negative key, the network is forced to actively maximize the feature distance between its final prediction and the initial estimate.

Let $\phi_i(\cdot)$ denote the VGG-19 features. The SGCR loss combines a relative contrastive term $\mathcal{R}_i$ and an absolute perceptual term $\mathcal{D}_i$:
\begin{equation}
	\mathcal{L}_{S} = \sum_{i=1}^{N_{VGG}} w_i \Big[ (1 - \mu) \mathcal{R}_i + (1 + \mu) \mathcal{D}_i \Big].
	\label{eq:L_SGCR_cn} 
\end{equation}
The components $\mathcal{D}_i$ and $\mathcal{R}_i$ are defined as:
\begin{align}
	\mathcal{D}_i &= |\phi_i(U^{(T)}) - \phi_i(X_{gt})|_1, \label{eq:perceptual_dist} \\
	\mathcal{R}_i &= \frac{\mathcal{D}_i}{|\phi_i(U^{(T)}) - \phi_i(U^{(0)})|_1 + \epsilon}, \label{eq:contrastive_ratio}
\end{align}
where $\mathcal{D}_i$ quantifies the perceptual discrepancy between the output and the ground truth.
$\mathcal{R}_i$ acts as a contrastive ratio that penalizes similarity to the coarse input $U^{(0)}$.
The parameter $\mu$ is a dynamic coefficient that increases linearly with training epochs.
This implements a curriculum learning strategy: In the early stages, the network focuses on suppressing severe artifacts, while in later stages, it shifts focus to fine-grained detail matching.
The joint objective function, $\mathcal{L}_j$, used for end-to-end training is defined as:
\begin{equation}
	\mathcal{L}_j = \mathcal{L}_{r} + \lambda_{g} \mathcal{L}_{S},
\end{equation}
where $\lambda_{g}$ is a hyperparameter balancing the contribution of the self-guided regularization.

\section{EXPERIMENTS}
\label{sec:experiments}

% =====================================================================
% A. Data Specification
% =====================================================================
\subsection{\textcolor{subsectioncolor}{Data Specification}}

\subsubsection{\textcolor{subsectioncolor}{Synthetic Data}}
Our method was evaluated using a synthetic dataset of CT images featuring simulated metal artifacts. The generation process adhered to the simulation procedure outlined from \cite{b10}. We randomly selected 1,200 clean CT images from the publicly available DeepLesion dataset \cite{b25} and sourced 100 metal masks from \cite{b10}. For the training data, 1,000 of these CT images and 90 metal masks were used for synthesis. Subsequently, a test set comprising 2,000 images was generated from the remaining 200 CT images and 10 metal masks.

\subsubsection{\textcolor{subsectioncolor}{Clinical CBCT Data}}
{We evaluated our method on two clinical dental datasets. The first, provided by the First Affiliated Hospital of Nanchang University, contains 20 cases with 200 slices each. The sinograms ($1536\times1600$) cover 1600 projections over $360^\circ$. These images feature varied metal implants without other anatomical abnormalities, ensuring reliable assessment(No. IIT [2025] LLS No. 482-1).

The second dataset was acquired using a LargeV dental CBCT system (Fig. \ref{fig:device}(a)). As shown in Fig. \ref{fig:device}(b), the geometry is defined by a Source-to-Isocenter Distance (SOD) of 200 mm and a Source-to-Image Distance (SID) of 640 mm. Operating at 100 kVp and 8 mA, the system captured 645 projections using a $1092 \times 1256$ detector with a 0.119 mm pitch. The final volume is $620 \times 620 \times 420$ voxels with 0.25 mm resolution.

	\begin{figure}[h!]
	\centerline{\includegraphics[width=\columnwidth]{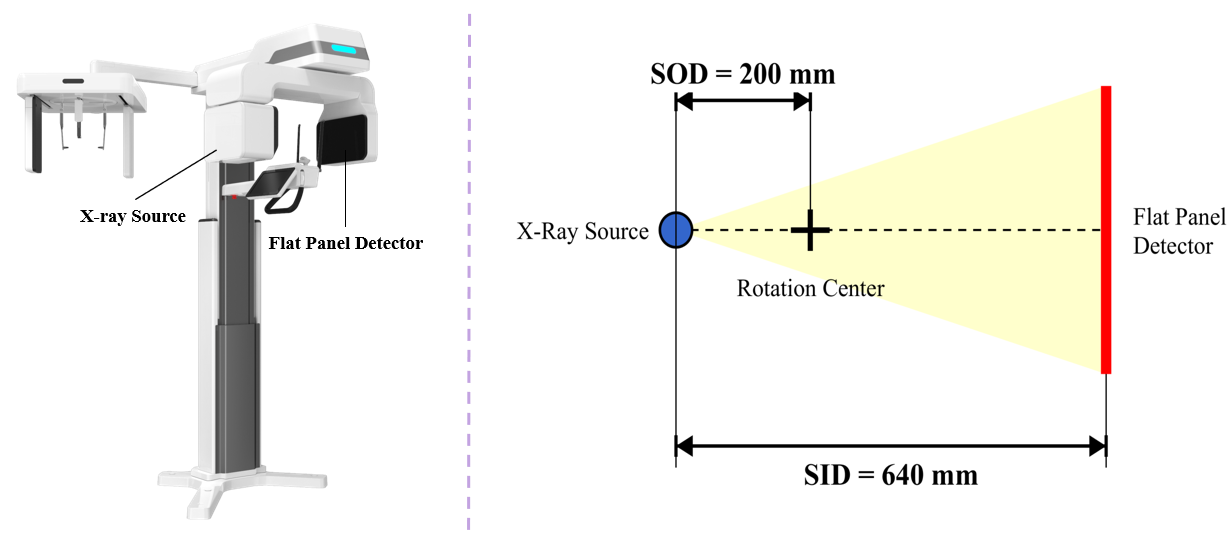}}
	\caption{Overview of the data acquisition system. (a) The Smart3D-X dental CBCT device equipped with a cone-beam X-ray generator and a flat panel detector. (b) Schematic diagram of the axial scanning geometry. The system features a source-to-object distance of 200 mm and a source-to-image distance of 640 mm. }
	\label{fig:device}
\end{figure}

}

\subsection{\textcolor{subsectioncolor}{Experimental Setup}}
{\normalcolor
	The proposed framework was implemented in Python using the PyTorch library. Model development and training were conducted on a personal workstation equipped with an NVIDIA RTX 5880 GPU. The network was trained for 200 epochs using the Adam optimizer with a batch size of 16. The learning rate was set to $5 \times 10^{-5}$, with $\beta_1=0.5$ and $\beta_2=0.999$. For the model architecture, the number of Mamba blocks per layer was configured as [1, 1, 2, 2, 4, 4, 2, 2, 1], and the number of iterations T was set to 2. To enhance training stability, data augmentation techniques, including random rotation and transposition, were applied. Model performance on the image reconstruction task was comprehensively evaluated using two standard image quality assessment metrics: The Peak Signal-to-Noise Ratio (PSNR) and the Structural Similarity Index (SSIM).

}

{\begin{table*}[t]
		\centering
		\caption{\textsc{Comparison of state-of-the-arts in terms of PSNR$\uparrow$ , SSIM$\uparrow$ under varying metal sizes on the synthesized DeepLesion dataset. $\uparrow$represents the higher the better. The bold font indicates the best values, respectively.}}
		\label{tab:quantitative_comparison_final}
		\setlength{\tabcolsep}{8pt}
		\begin{tabular}{lccccc|c}
			\toprule
			\textbf{Method} & \textbf{Large Metal} & $\xrightarrow{\hspace*{0.5cm}}$ & \textbf{Medium Metal} & $\xrightarrow{\hspace*{0.5cm}}$ & \textbf{Small Metal} & \textbf{Average} \\
			\midrule
			LI\cite{b4}           & 30.17 / 0.7513 & 31.24 / 0.7687 & 32.48 / 0.8106 & 33.56 / 0.8329 & 34.14 / 0.8497 & 32.32 / 0.8026 \\
			NMAR\cite{b5}         & 31.23 / 0.7824 & 32.19 / 0.8038 & 33.42 / 0.8461 & 34.68 / 0.8645 & 35.33 / 0.8814 & 33.37 / 0.8356 \\
			ADN\cite{b29}         & 34.52 / 0.8513 & 35.38 / 0.8695 & 36.21 / 0.8927 & 37.06 / 0.9064 & 37.84 / 0.9152 & 36.20 / 0.8870 \\
			CNNMAR\cite{b10}      & 35.14 / 0.9423 & 35.82 / 0.9538 & 36.53 / 0.9654 & 37.19 / 0.9706 & 37.83 / 0.9752 & 36.50 / 0.9615 \\
			InDuDoNet\cite{b18}   & 37.23 / 0.9614 & 37.91 / 0.9668 & 38.56 / 0.9755 & 39.24 / 0.9796 & 39.92 / 0.9813 & 38.57 / 0.9729 \\
			InDuDoNet+\cite{b32}  & 37.54 / 0.9632 & 38.18 / 0.9708 & 38.86 / 0.9785 & 39.45 / 0.9824 & 40.13 / 0.9854 & 38.83 / 0.9761 \\
			ACDNet\cite{b26}      & 42.13 / 0.9783 & 43.17 / 0.9839 & 44.19 / 0.9894 & 45.23 / 0.9912 & 46.21 / 0.9934 & 44.19 / 0.9872 \\
			MAIL\cite{b27}        & 41.56 / 0.9654 & 42.42 / 0.9703 & 43.31 / 0.9757 & 44.18 / 0.9795 & 45.14 / 0.9823 & 43.32 / 0.9746 \\
			TransCT\cite{b38}     & 41.22 / 0.9613 & 42.15 / 0.9676 & 43.08 / 0.9742 & 43.91 / 0.9788 & 44.83 / 0.9804 & 43.04 / 0.9725 \\
			\midrule
			\textbf{AS-Mamba}     & \textbf{43.53} / \textbf{0.9853} & \textbf{44.29} / \textbf{0.9888} & \textbf{45.06} / \textbf{0.9934} & \textbf{45.78} / \textbf{0.9952} & \textbf{46.54} / \textbf{0.9963} & \textbf{45.04} / \textbf{0.9918} \\
			\bottomrule
		\end{tabular}
\end{table*}

\begin{figure*}[!t]
	\centerline{\includegraphics[width=\textwidth]{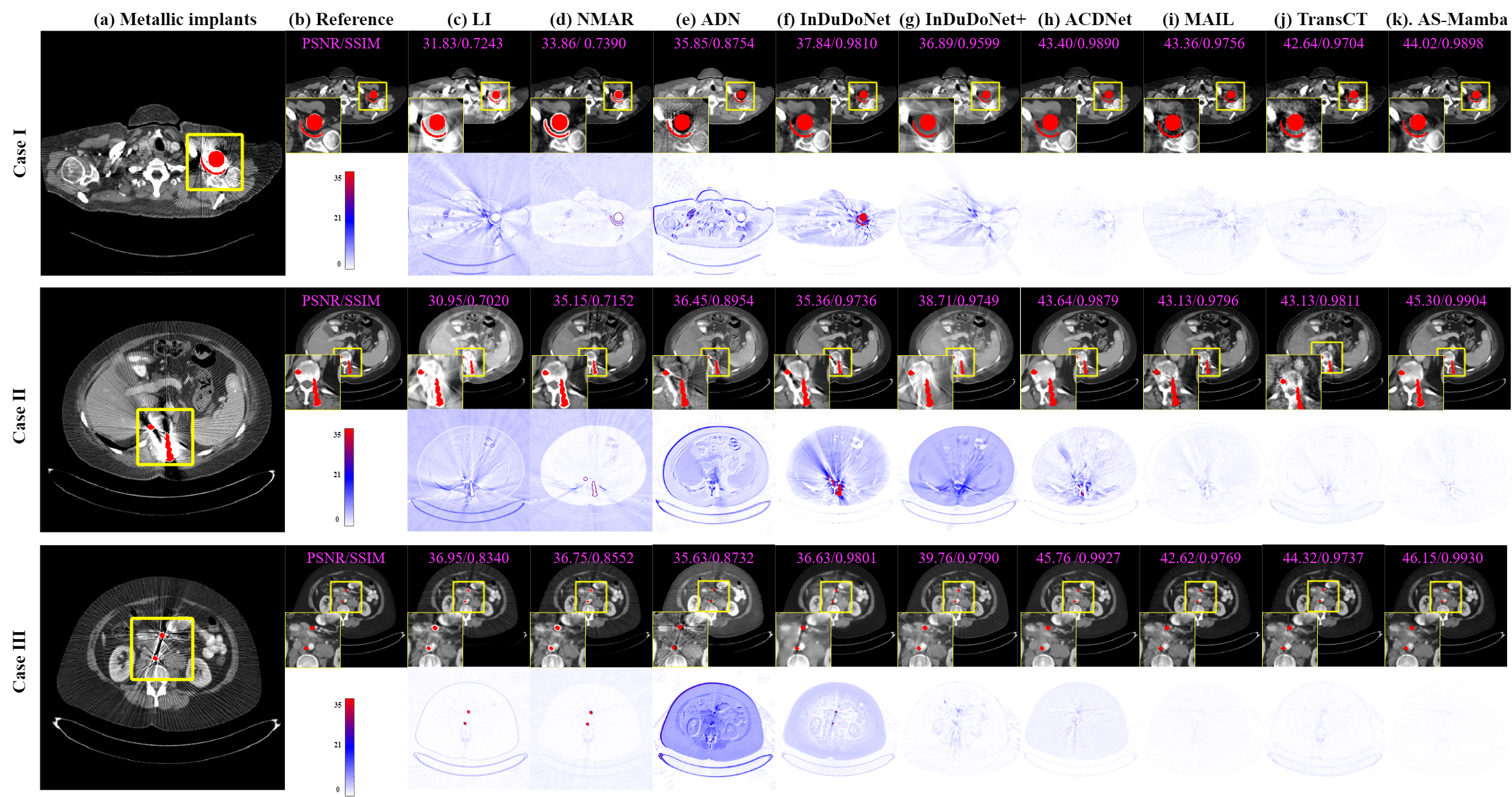}}
	\caption{Visual comparison of MAR results by different methods on three synthetic images. The metal-corrupted CT images are randomly selected from the test dataset. The metal masks are highlighted in red for better visualization. The display window is [-175,275] HU. The image part within the yellow box is the region of interest (ROI).}
	\label{fig4}
	\vspace{-0.5cm}
\end{figure*}

}

\subsection{\textcolor{subsectioncolor}{Experimental Results on DeepLesion Data}}
\normalcolor
\subsubsection{Quantitative Comparisons With State-of-the-Arts}
In order to assess the effectiveness of the proposed AS-Mamba, we evaluated the performance and compared with popular or state-of-the-arts, including LI \cite{b4}, NMAR\cite{b5}, CNNMAR\cite{b10}, ADN\cite{b29}, ACDNet\cite{b26}, InDuDoNet\cite{b18}, InDuDoNet+\cite{b32}, MAIL\cite{b27}, and TransCT\cite{b38}.

Table~\ref{tab:quantitative_comparison_final} presents the comparative results in terms of PSNR and SSIM. As observed, prior-based traditional methods like NMAR outperform simple LI, while deep learning-based approaches, such as ACDNet and MAIL, achieve significantly higher scores, demonstrating the superiority of data-driven paradigms. Among the competing deep methods, our proposed framework achieves the highest performance, yielding a PSNR of 45.04 dB and an SSIM of 0.9918, which represents a substantial improvement over the second-best performing methods, ACDNet and MAIL.

\begin{figure}[h!]
	\centerline{\includegraphics[width=0.9\columnwidth]{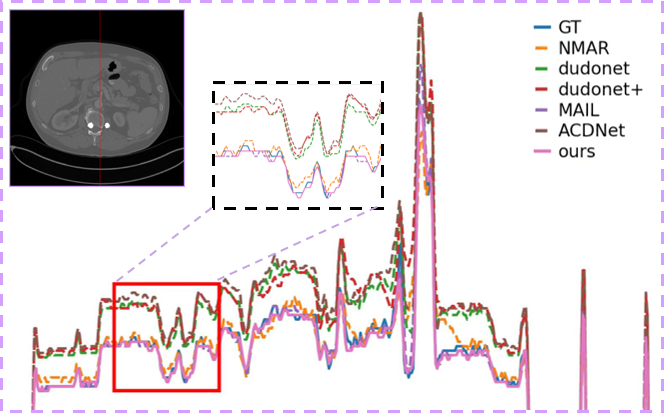}}
	\caption{The intensity profiles along the specified red line of a randomly selected image after applying different MAR methods.}
	\label{fig5}
\end{figure}

\subsubsection {Intensity Profile Analysis}
To further evaluate pixel-level fidelity and edge preservation, we conducted a 1D intensity profile analysis. As shown in Fig. \ref{fig5}, we plotted the intensity distribution along the red line on a representative slice containing complex tissues. The comparison reveals clear differences between methods. The orange dashed curve of NMAR\cite{b5} shows significant deviations from the blue solid Ground Truth line, indicating severe intensity distortion. While deep learning methods improve the general trend, they still struggle with local details. Specifically, in the Region of Interest, the brown dashed profile of ACDNet and the green dashed profile of InDuDoNet\cite{b18} tend to produce unstable peaks, reflecting a loss of contrast accuracy. Even the competitive MAIL method, shown as the purple dashed line, exhibits slight misalignment with the GT in sharp transition areas. In contrast, the pink solid line of our method tightly follows the GT curve throughout the scanning path. This precise alignment confirms that AS-Mamba offers superior performance in preserving anatomical structures and suppressing artifacts compared to competing approaches.

\subsubsection{Qualitative Analysis}
We present a visual comparison of different MAR methods on simulated DeepLesion data in Fig. \ref{fig4}, featuring three distinct implant sizes with overlaid metal masks and magnified ROIs. The original images exhibit severe dark band artifacts between the implants. LI, NMAR and even the deep learning-based MAIL, while reducing conspicuous streaks, often introduce secondary artifacts and loss of detail due to sinogram discontinuities. While deep learning approaches generally outperform traditional ones, methods like InDuDoNet+ tend to produce overly smooth results lacking geometric constraints. For larger implants, ACDNet and MAIL improve upon ADN but still leave residual artifacts. In contrast, our proposed method effectively suppresses the vast majority of artifacts while accurately preserving fine anatomical details.

\subsection{Performance on Clinical CBCT Data}
To validate clinical applicability, we evaluated AS-Mamba on real-world dental CBCT scans. Since ground truth is unavailable for clinical data, we employed three objective metrics: Standard Deviation (STD) to measure tissue smoothness, Contrast-to-Noise Ratio (CNR) to assess structural clarity, and a subjective Clinical Score.

The Clinical Score was derived from a blind observer study conducted by medical experts at the First Affiliated Hospital of Nanchang University. Specifically, twelve dentists with intermediate professional titles from this institution independently assessed the reconstructed images. The evaluation followed a 5-point scale, focusing on artifact suppression, tissue visibility, and overall diagnostic clarity. The scoring criteria were strictly defined as follows:
 
1: Severe artifacts rendering the image unusable; 

2: Significant artifacts obscuring key structures; 

3: Moderate artifacts but sufficient for general diagnosis; 

4: Minor artifacts with clear anatomical details; 

5: Negligible artifacts with optimal diagnostic quality. 

The final Clinical Score represents the average of these independent ratings.

The visual comparison on the LargeV dataset is presented in Fig. 8, where AS-Mamba demonstrates its superior capability in tracing and suppressing directional metal streaks via the Mamba scan mechanism, while avoiding the blurring common in traditional methods or the over-smoothing seen in existing deep learning models. To further assess clinical utility, Fig. 9 and Table \ref{tab:clinical_metrics} provide the evaluation on real-world dental CBCT scans from the First Affiliated Hospital of Nanchang University. AS-Mamba achieves the best quantitative results with the lowest STD of $31.25$ HU and the highest CNR of 0.98, indicating effective artifact removal and contrast recovery. Most importantly, our method received the highest Clinical Score of 4.00 from  specialists, confirming its ability to provide the most reliable anatomical details for clinical diagnosis.

\begin{table}[h!] 
	\centering
	\caption{\textsc{Quantitative evaluation on clinical dental CBCT data. $\uparrow$ indicates higher is better, and $\downarrow$ indicates lower is better. The bold font indicates the best values.}}
	\label{tab:clinical_metrics}
	\renewcommand{\arraystretch}{1.3} 
	\setlength{\tabcolsep}{2.5pt}  % 设置列间空白为2.5pt，模仿第一个表格的紧凑性
	\begin{tabular}{lccc}  % 使用普通tabular，列对齐为lccc
		\toprule
		\textbf{Method} & \textbf{STD (HU) $\downarrow$} & \textbf{CNR $\uparrow$} & \textbf{Clinical Score $\uparrow$} \\
		\midrule
		LI\cite{b4}        & 48.94 & 0.60 & 2.45 \\
		NMAR\cite{b5}     & 44.64 & 0.65 & 2.80 \\
		MAIL\cite{b27}     & 31.48 & 0.78 & 3.60 \\
		ADN\cite{b29}      & 36.82 & 0.81 & 3.45 \\ 
		ACDNet\cite{b26}   & 33.69 & 0.88 & 3.85 \\
		\midrule
		\textbf{AS-Mamba} & \textbf{31.25} & \textbf{0.98} & \textbf{4.00} \\
		\bottomrule
	\end{tabular}
\end{table}
\begin{figure*}[t]
	\centerline{\includegraphics[width=\textwidth]{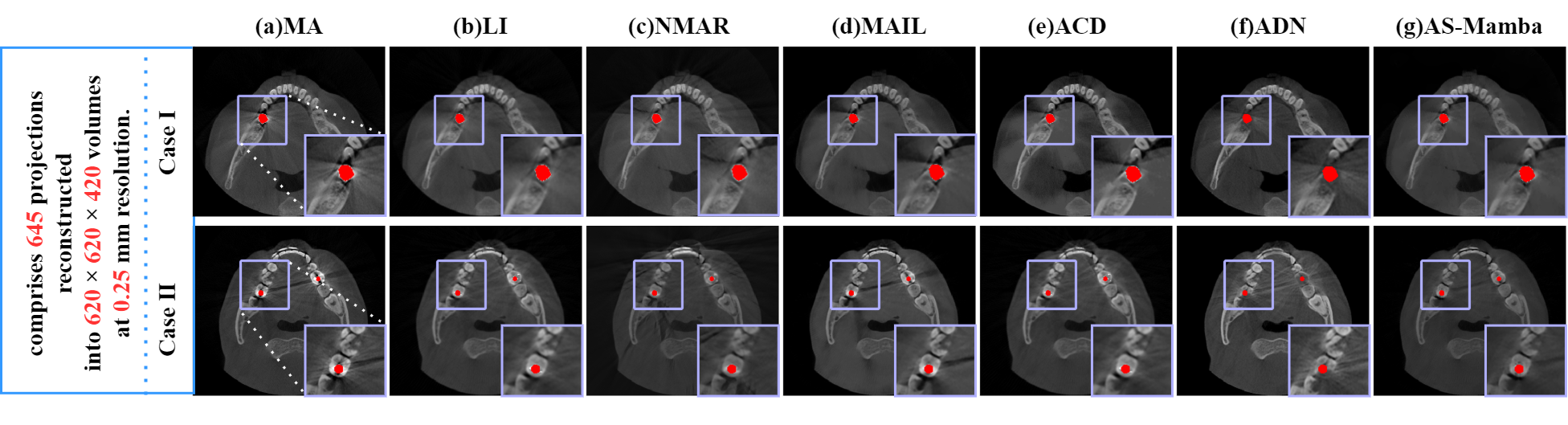}}
	\caption{Comparison of different MAR methods on the LargeV clinical dataset of patients undergoing oral and maxillofacial examinations with metallic structures of various sizes. The magnified region of interest (ROI) are highlighted in purple boxes.}
	\label{fig6}
	\vspace{-0.5cm}
\end{figure*}

\begin{figure*}[t]
	\centerline{\includegraphics[width=\textwidth]{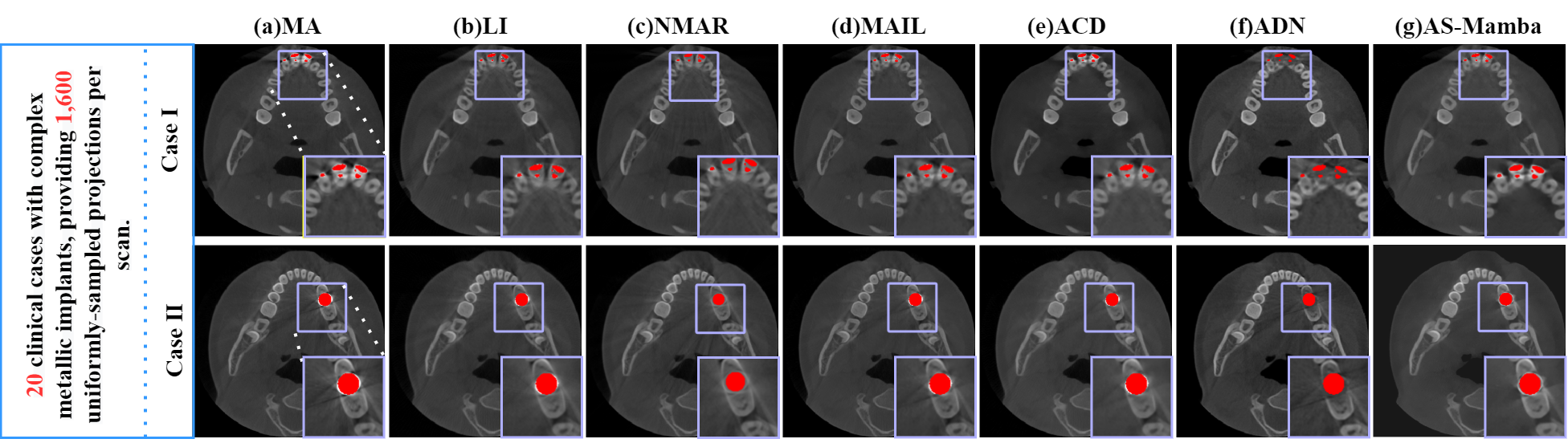}}
	\caption{Visual results of different MAR methods on the Stomatology Center of the First Affiliated Hospital of Nanchang University dental dataset with metallic implants. The magnified region of interest (ROI) are highlighted in purple boxes.}
	\label{fig7}
	\vspace{-0.5cm}
\end{figure*}

\subsection{\textcolor{subsectioncolor}{Ablation Study}}

To validate AS-Mamba, we conducted ablation studies on the synthetic dataset to systematically evaluate the hybrid-domain framework, asymmetric processors, iterative refinement, and SGCR loss. All models were trained and evaluated under identical settings.

\subsubsection{Effectiveness of Frequency Decoupling and Iterative Refinement}

This study aims to verify the superiority of the proposed framework. We conducted ablation experiments to evaluate the individual contributions of wavelet‑domain processing, frequency‑decoupling strategy, and iterative refinement to the performance of the AS‑Mamba model. We have designed the following ablation experiments:

	Frequency-Decoupled Reconstruction Network (FDR-Net): This model performs the wavelet decomposition and dual-branch processing but omits the subsequent iterative refinement. The output from the initial reconstruction is the final result.
	
	Spatial-Domain Iterative Network (SDI-Net): This model removes the entire wavelet-domain processing pipeline. It directly processes the concatenated input with a U-Net, followed by the iterative refinement stages.
	
	Unified Wavelet Processing Network (UWP-Net): This model performs the wavelet transform but does not decouple the coefficients into low and high frequencies. Unified network processes the wavelet coefficients together before the inverse transform.

% 【图 8】
\begin{figure}[h!]
	\centerline{\includegraphics[width=\linewidth]{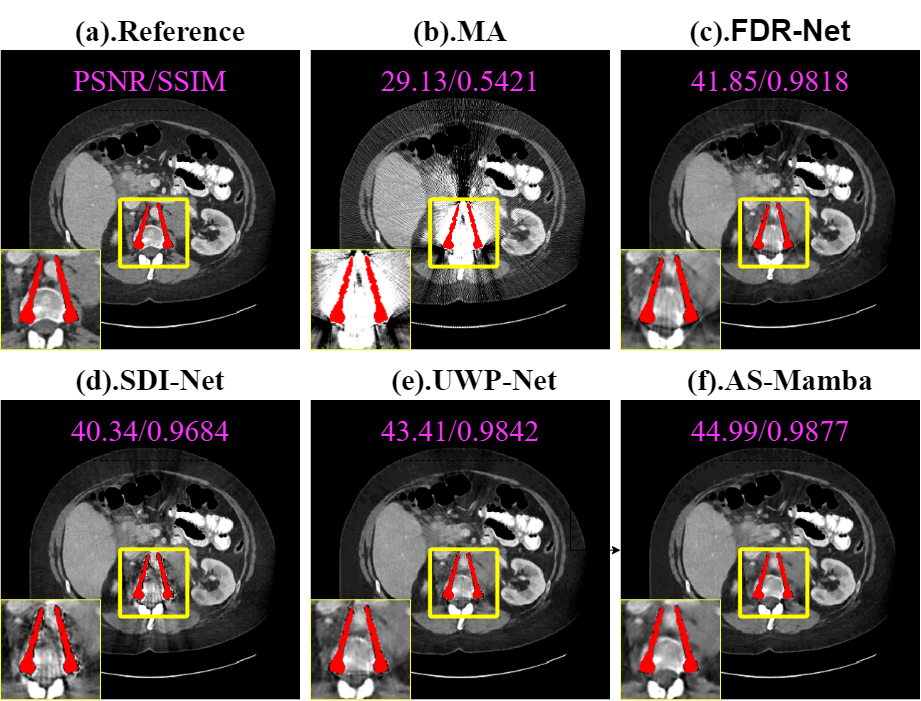}}
	\caption{Visual comparison for the framework component analysis. The display window is [-175,275] HU. The image part within the yellow box is the region of interest.}
	\label{fig8}
	\vspace{-0.5cm}
\end{figure}

% 【表格 II】
\begin{table}[h!]
	\centering
	\caption{\textsc{Effect of core framework components on the synthesized dataset. The column ``Average'' represents the PSNR/SSIM averagely computed across different metal sizes.}}
	\label{tab:ablation_study}
	\renewcommand{\arraystretch}{1.3}
	\setlength{\tabcolsep}{2.5pt}
	\begin{tabular}{lccc|c}
		\toprule
		\textbf{Method} & \textbf{Large} & \textbf{Medium} & \textbf{Small} & \textbf{Average} \\
		\midrule
		FDR-Net   & 40.53 / 0.9754 & 42.84 / 0.9825 & 44.13 / 0.9864 & 42.50 / 0.9814 \\
		SDI-Net   & 38.54 / 0.9513 & 40.65 / 0.9634 & 42.14 / 0.9712 & 40.44 / 0.9620 \\
		UWP-Net   & 41.83 / 0.9792 & 43.94 / 0.9863 & 45.23 / 0.9894 & 43.67 / 0.9850 \\
		\midrule
		\textbf{AS-Mamba} & \textbf{43.24 / 0.9814} & \textbf{45.12 / 0.9873} & \textbf{46.64 / 0.9912} & \textbf{44.97 / 0.9866} \\
		\bottomrule
	\end{tabular}
\end{table}

Quantitative results in Table~\ref{tab:ablation_study} show that AS-Mamba consistently outperforms all variants. Specifically, SDI-Net exhibits the lowest metrics, confirming that pure spatial methods struggle to separate spatially intertwined artifacts and tissues. The superiority over UWP-Net validates the necessity of frequency decoupling, as uniform processing fails to effectively target high-frequency streaks. Furthermore, FDR-Net falls short of the full model, indicating that image-domain iterative refinement is essential for enforcing global consistency. Visual comparisons in Fig. \ref{fig8} confirm these findings, where AS-Mamba achieves the sharpest restoration compared to the blurred or artifact-prone baselines.

\subsubsection{Validation of the Asymmetric Frequency Processors}
To validate the branch-specific designs, we replaced the specialized modules with standard convolutional networks: Variant A substitutes the Fourier-enhanced DEN in the low-frequency branch, and Variant B substitutes the Mamba-based module in the high-frequency branch.

% 【表格 III】
\begin{table}[h!]
	\centering
	\caption{\textsc{Effect of different variants on the DeepLesion dataset. The column ``Average'' denotes the PSNR/SSIM results averagely computed on the synthesized DeepLesion dataset.}}
	\label{tab:ablation_variants}
	\renewcommand{\arraystretch}{1.3} 
	\setlength{\tabcolsep}{2.5pt}  
	\begin{tabular}{lccc|c}
		\toprule
		\textbf{Method} & \textbf{Large} & \textbf{Medium} & \textbf{Small} & \textbf{Average} \\ 
		\midrule
		Variant A & 41.24 / 0.9754 & 43.63 / 0.9835 & 45.69 / 0.9883 & 43.52 / 0.9824 \\
		Variant B & 40.23 / 0.9723 & 42.54 / 0.9814 & 44.56 / 0.9867 & 42.44 / 0.9801 \\
		\midrule
		\textbf{AS-Mamba} & \textbf{43.24 / 0.9814} & \textbf{45.12 / 0.9873} & \textbf{46.64 / 0.9912} & \textbf{44.97 / 0.9866} \\
		\bottomrule
	\end{tabular}
\end{table}

As shown in Table~\ref{tab:ablation_variants}, both modifications degrade performance. Specifically, Variant A causes a notable decline in SSIM, confirming that the Fourier-enhanced design is vital for preserving global structural integrity. In contrast, Variant B results in a larger drop in PSNR. This highlights the inability of standard convolutions to handle non-local irregularities, demonstrating the superiority of Mamba in modeling and suppressing the long-range streak artifacts inherent to high-frequency components.

\subsubsection{Effectiveness of the Self-Guided Contrastive Regularization}
We further conduct an ablation study on contrastive regularization, with the experimental results presented in Table~\ref{tab:ablation_contrastive}. It can be observed that the conventional contrastive paradigm provides limited performance gains. In comparison, the proposed self-guided contrastive regularization enhances the model’s feature discriminability by treating the coarse restored image as informative guidance, thereby leading to improved dehazing results.

\begin{table}[h!] 
	\centering
	\caption{\textsc{Ablation Study on Contrastive Regularization}}
	\label{tab:ablation_contrastive} 
	\renewcommand{\arraystretch}{1.3} 
	\setlength{\tabcolsep}{2.5pt} 
	\begin{tabular}{lcccc}
		\hline
		& w/o CR & CR & SGCR & CR+SGCR \\
		\hline
		PSNR & 42.41 & 42.63 & 42.73 & \textbf{43.97} \\
		SSIM & 0.9786 & 0.9805 & 0.9805 & \textbf{0.9853} \\
		\hline
	\end{tabular}
\end{table}

\section{Conclusion}
In this study, we proposed AS-Mamba, a novel network designed to address the complex challenge of metal artifact reduction. Recognizing that artifacts manifest as both directional streaks and global shading, we developed an asymmetric dual-branch architecture. The high-frequency branch leverages the linear complexity and long-range modeling capability of the Mamba model to effectively trace and remove continuous streaks. Complementarily, the low-frequency branch utilizes Fourier transform to rectify global intensity inhomogeneity caused by beam hardening. Furthermore, to bridge the domain gap between synthetic training data and real-world clinical scenarios, we introduced a Self-Guided Contrastive Regularization (SGCR) strategy.

Comprehensive experiments on both synthesized and clinical dental CBCT datasets demonstrate that AS-Mamba significantly outperforms state-of-the-art methods in preserving anatomical details and suppressing artifacts. Despite these advancements, our current approach relies on supervised learning with paired data, which may limit its flexibility in some clinical settings. Future work will focus on developing unsupervised or semi-supervised adaptations to better utilize unlabeled clinical data and extending the framework to 3D volumetric reconstruction.

\end{document}